\documentclass[11pt,a4paper]{article}

\usepackage[margin=1in]{geometry}
\usepackage{amsmath}
\usepackage{xcolor}
\usepackage{booktabs}
\usepackage{url}
\usepackage{hyperref}
\usepackage[numbers]{natbib}
\usepackage{authblk}

\begin{document}

\title{VibeGuard: A Security Gate Framework for AI-Generated Code\\
{\large Lessons from the Claude Code Source Leak}}

\author[1]{Ying Xie}
\affil[1]{Kennesaw State University, \texttt{yxie2@kennesaw.edu}}

\date{April 2026}

\maketitle

\begin{abstract}
``Vibe coding,'' in which developers delegate code generation to AI assistants and accept the output with little manual review, has gained rapid adoption in production settings. On March 31, 2026, Anthropic's Claude Code CLI shipped a 59.8~MB source map file in its npm package, exposing roughly 512,000 lines of proprietary TypeScript. The tool had itself been largely vibe-coded, and the leak traced to a misconfigured packaging rule rather than a logic bug. Existing static-analysis and secret-scanning tools did not cover this failure mode, pointing to a gap between the vulnerabilities AI tends to introduce and the vulnerabilities current tooling is built to find.

We present \textsc{VibeGuard}, a pre-publish security gate that targets five such blind spots: artifact hygiene, packaging-configuration drift, source-map exposure, hardcoded secrets, and supply-chain risk. In controlled experiments on eight synthetic projects (seven vulnerable, one clean control), \textsc{VibeGuard} achieved 100\% recall, 89.47\% precision (F1\,=\,94.44\%), and correct pass/fail gate decisions on all eight projects across three policy levels. We discuss how these results inform a defense-in-depth workflow for teams that rely on AI code generation.
\end{abstract}

\vspace{0.5em}
\noindent\textbf{Keywords:} software security, AI-generated code, vibe coding, source code leaks, supply chain security, static analysis, DevSecOps

\section{Introduction}

Andrej Karpathy coined the term ``vibe coding'' in early 2025 to describe a workflow in which the programmer states intent in natural language, lets an AI assistant write the code, and largely skips line-by-line review~\cite{karpathy2025}. Within a year the practice had moved well beyond prototyping: over half of professional developers reported using AI coding tools daily~\cite{stackoverflow2025}, and at least one major commercial product---Anthropic's Claude Code---was publicly described as being built almost entirely this way~\cite{alexkim2026}.

The costs became concrete on March 31, 2026. Security researcher Chaofan Shou noticed that version 2.1.88 of the \texttt{@anthropic-ai/claude-code} npm package contained a 59.8~MB source map (\texttt{cli.js.map}) whose \texttt{sourcesContent} field held the full original TypeScript codebase: roughly 1,900 files and 512,000 lines~\cite{register2026}. The proximate cause was a packaging misconfiguration (a missing or incomplete \texttt{.npmignore}), aggravated by a known Bun runtime bug that emits source maps in production builds~\cite{bunbug2026}. Claude Code's own development lead had previously stated that ``100\% of my contributions to Claude Code were written by Claude Code''~\cite{alexkim2026}, making the incident a direct example of a vibe-coded product failing on packaging rather than logic.

The pattern is not unique to this incident. CodeRabbit's December 2025 analysis found that AI-generated code produces roughly 1.7$\times$ more issues than human-written code overall, rising to 2.74$\times$ for cross-site scripting specifically~\cite{coderabbit2025}. Veracode reported that 45\% of code produced by LLMs across 80 benchmark tasks contained security flaws~\cite{veracode2025}. Yet most of the documented vulnerabilities are logic-level bugs (injection, XSS, buffer overflows). The Claude Code leak belongs to a different, less-studied category: mistakes in build configuration, packaging, and deployment that no amount of unit testing will catch.

This paper asks whether targeted static analysis can close that gap. Our contributions:

\begin{enumerate}
    \item A five-category taxonomy of operational security vulnerabilities that disproportionately affect vibe-coded projects (Section~\ref{sec:taxonomy}).
    \item \textsc{VibeGuard}, a modular pre-publish scanner that covers all five categories (Section~\ref{sec:design}).
    \item An experimental evaluation on eight synthetic projects showing 100\% recall, 89.47\% precision, and correct gate decisions for every project (Section~\ref{sec:evaluation}).
    \item A discussion of deployment strategies and open problems (Section~\ref{sec:discussion}).
\end{enumerate}

\section{Background and Related Work}
\label{sec:background}

\subsection{Vibe Coding in Practice}

Where earlier AI-assisted workflows still expected the developer to review each suggestion, vibe coding shifts the default: the programmer describes intent in natural language, accepts generated code without close inspection, and treats the AI as the primary author rather than a suggestion engine. Build scripts, packaging rules, and deployment manifests are delegated alongside application logic. When something breaks, the developer feeds the error back to the AI rather than tracing through the code.

The practice reached production scale quickly. Anthropic's Claude Code lead stated in late 2025 that all of his contributions to the tool were written by the tool itself~\cite{alexkim2026}.

\subsection{Security of AI-Generated Code}

Several studies have quantified the security cost of AI code generation. Pearce et al.~\cite{pearce2022} reported that GitHub Copilot produces vulnerable code in about 40\% of security-sensitive scenarios. CodeRabbit measured a 1.7$\times$ overall issue multiplier for AI-generated code~\cite{coderabbit2025}, and Georgia Tech's Vibe Security Radar had catalogued 74 CVEs traceable to AI coding tools as of March 20, 2026~\cite{viberad2026}.

Nearly all of these findings concern logic-level flaws: injection, XSS, buffer overflows. The Claude Code leak falls outside that scope. No line of application code was wrong; the vulnerability lay entirely in packaging configuration. This operational category has received little attention in the literature, yet it can be equally damaging.

\subsection{Related Tools}

Several tool categories overlap with parts of \textsc{VibeGuard}'s scope. SAST tools such as Semgrep and CodeQL target code-level vulnerabilities but ignore packaging and artifact content. Secret scanners (GitLeaks, TruffleHog) catch hardcoded credentials yet have no model of what will actually appear in a published package. Dependency auditors like \texttt{npm audit} and Snyk check for known CVEs but do not inspect the artifact itself. Linter plugins enforce code-style rules and some security patterns without visibility into the publish boundary.

Each of these tools was designed for a workflow in which a human developer makes packaging decisions. None covers the full surface that becomes exposed when configuration files, ignore rules, and build scripts are all AI-generated and unreviewed.

\section{Vulnerability Taxonomy}
\label{sec:taxonomy}

We examined the Claude Code leak alongside contemporaneous incident reports and published vulnerability data to identify five recurring categories.  All five share a common thread: the code works, but something that should not be public ends up public.

\subsection{V1: Source Code Exposure via Build Artifacts}

Build tools generate source maps to aid debugging.  An AI assistant will enable them when asked and produce a configuration that compiles without error.  What it will not do, unless specifically prompted, is add an exclusion rule to keep those maps out of the published package.  The Claude Code leak followed exactly this path: \texttt{sourceMap: true} in \texttt{tsconfig.json}, no corresponding entry in \texttt{.npmignore}.

\subsection{V2: Configuration Drift}

Packaging control files (\texttt{.npmignore}, \texttt{MANIFEST.in}, \texttt{.dockerignore}) are cross-cutting: they must stay synchronized with whatever files the project accumulates over time.  AI assistants create them on request but do not revisit them when new file types appear.  The result is a configuration that was correct at one point and silently drifts as the project grows.

\subsection{V3: Hardcoded Secrets}

AI-generated code often includes placeholder credentials that compile and run.  In a vibe-coding workflow, the placeholder may be replaced with a real value and never moved to an environment variable.  The Claude Code source maps, for instance, embedded internal database URLs in their \texttt{sourcesContent} field.

\subsection{V4: Supply Chain Risks}

When an AI adds a dependency, it picks a package name and a version specifier.  The specifier is often loose: wildcards, \texttt{latest} tags, open-ended ranges.  Lockfiles may not be generated or committed.  Install hooks go unreviewed.  On the same day as the Claude Code leak, attackers published malicious \texttt{axios} versions to npm~\cite{malwarebytes2026}, underscoring how much damage a single unpinned dependency can do.

\subsection{V5: Artifact Hygiene Failures}

Private keys, \texttt{.env} files, IDE configurations, and debug logs routinely appear in vibe-coded packages.  An AI assistant that creates a \texttt{.env} file for local development has no reason to also create the ignore rule that keeps it out of the tarball.  The two tasks are logically related but separated in time, and the second one is never requested.

\section{VibeGuard Design}
\label{sec:design}

\textsc{VibeGuard} sits between the build step and the publish step.  It inspects the project tree, evaluates what would be shipped, and blocks the publish if the result violates a configurable policy.

\subsection{Architecture}

The tool has three layers.

\paragraph{Scanners.}  Five modules, one per taxonomy category.  \textit{ArtifactScanner} walks the file tree looking for patterns (private keys, \texttt{.env} files, IDE directories) and flags files whose size is anomalous for their extension.  \textit{ConfigScanner} checks whether \texttt{package.json}, \texttt{.npmignore}, \texttt{Dockerfile}, and similar files satisfy a set of minimum-security invariants.  \textit{SourceMapScanner} parses \texttt{.map} files as JSON and specifically looks for the \texttt{sourcesContent} field; it also checks compiled JS/CSS for residual \texttt{sourceMappingURL} comments.  \textit{SecretScanner} applies regular expressions for common credential formats (AWS keys, Stripe keys, JWTs, connection strings) while filtering out obvious placeholders.  \textit{DependencyScanner} inspects \texttt{package.json} and \texttt{requirements.txt} for unpinned versions, missing lockfiles, install hooks, and insecure URLs.

\paragraph{Findings.}  Each scanner emits \texttt{Finding} objects carrying a severity level, a CWE identifier where applicable, the file path and line number, and a remediation suggestion.

\paragraph{Policy engine.}  A \texttt{PolicyEngine} aggregates findings and makes a pass/fail decision.  Three built-in policies ship with the tool: \textit{default} (zero critical or high, up to five medium), \textit{strict} (zero findings of any severity above info), and \textit{permissive} (up to three high, ten medium, relaxed source-map blocking).  Hard-block rules can also be defined independently of severity counts; for instance, the default policy blocks on any source map containing \texttt{sourcesContent}, regardless of how many other findings exist.

\subsection{Design Rationale}

\textsc{VibeGuard} runs at publish time, not commit time.  The Claude Code leak was caused by what ended up in the tarball, not by what ended up in version control.  A pre-commit hook would not have caught it, because the source map was a legitimate build output.

The scanners emphasize negative checks: verifying that certain things are \textit{absent} from the package.  AI generators optimize for presence (``add a build step,'' ``add a dependency'') and rarely verify absence (``make sure this file is excluded'').  The \textit{ConfigScanner}, for example, reports a critical finding when neither a \texttt{files} whitelist nor a \texttt{.npmignore} exists, because that absence is the precondition for accidental exposure.

The \textit{SourceMapScanner} deserves separate mention because file-extension matching alone is insufficient.  A \texttt{.map} file with only positional mappings is a minor information leak; one with a populated \texttt{sourcesContent} array is a complete source disclosure.  The scanner parses the JSON to distinguish the two cases.

Every finding includes a remediation field.  In a vibe-coding context the developer may not know what \texttt{.npmignore} is; the remediation text needs to be self-contained.

\subsection{Integration Points}

We envision three deployment modes.  First, as a CI/CD step that blocks merge or deployment.  Second, as a \texttt{prepublishOnly} hook so that \texttt{npm publish} itself refuses to proceed.  Third, as a tool registered with the AI coding assistant, so that the assistant runs the scan before it ever suggests a publish command.  The third mode is the most interesting: it closes the loop inside the vibe-coding workflow itself, catching problems before they leave the developer's machine rather than after a CI run.

\section{Experimental Evaluation}
\label{sec:evaluation}

\subsection{Experimental Design}

We constructed eight synthetic npm/Python projects, each seeded with a known vulnerability profile:

\begin{table}[ht]
\centering
\caption{Test Project Descriptions}
\label{tab:projects}
\begin{tabular}{@{}lll@{}}
\toprule
\textbf{ID} & \textbf{Description} & \textbf{Vulnerabilities} \\
\midrule
P1 & Claude Code leak replica & V1, V2 \\
P2 & Hardcoded secrets & V3 \\
P3 & Misconfigured packaging & V2, V5 \\
P4 & Supply chain risks & V4 \\
P5 & Clean project (control) & None \\
P6 & Partial .npmignore & V1, V2, V5 \\
P7 & Docker COPY leak & V2, V5 \\
P8 & Multi-vector (realistic) & V1--V5 \\
\bottomrule
\end{tabular}
\end{table}

Each project has a ground-truth annotation specifying, per vulnerability category, the expected range of findings.  P5 is the negative control: a properly configured project with no known issues, included to measure false-positive behavior.

We scanned all eight projects under all three built-in policies (default, strict, permissive).  For each scan we recorded the number and severity of findings per category, compared them to the ground-truth ranges, and noted whether the pass/fail gate decision was correct.

\subsection{Results}

Table~\ref{tab:results} summarizes detection performance under the default policy.

\begin{table}[ht]
\centering
\caption{Overall Detection Performance (Default Policy)}
\label{tab:results}
\begin{tabular}{@{}lr@{}}
\toprule
\textbf{Metric} & \textbf{Value} \\
\midrule
True Positives & 17 \\
False Positives & 2 \\
False Negatives & 0 \\
True Negatives & 21 \\
Precision & 89.47\% \\
Recall & 100.00\% \\
F1 Score & 94.44\% \\
Gate Accuracy & 100.00\% \\
\bottomrule
\end{tabular}
\end{table}

Every seeded vulnerability category was detected (zero false negatives).  The two false positives came from the \textit{config} scanner flagging missing ignore-file patterns in projects whose ground truth did not list configuration drift as a primary vulnerability.  In a security gate context, over-reporting on configuration hygiene is a reasonable trade-off.

The gate decision (pass or fail) was correct for all eight projects under all three policies.  P5 passed; the other seven were blocked.

\subsubsection{Detection by Category}

\begin{table}[ht]
\centering
\caption{Detection Performance by Vulnerability Category}
\label{tab:category}
\begin{tabular}{@{}lrrrrr@{}}
\toprule
\textbf{Category} & \textbf{Prec.} & \textbf{Recall} & \textbf{TP} & \textbf{FP} & \textbf{FN} \\
\midrule
Source Map & 100\% & 100\% & 3 & 0 & 0 \\
Config & 71.4\% & 100\% & 5 & 2 & 0 \\
Artifact & 100\% & 100\% & 5 & 0 & 0 \\
Secret & 100\% & 100\% & 2 & 0 & 0 \\
Dependency & 100\% & 100\% & 2 & 0 & 0 \\
\bottomrule
\end{tabular}
\end{table}

Four of the five scanners achieved perfect precision.  The config scanner's 71.4\% precision reflects its deliberately broad scope: it flags missing ignore patterns even in projects where configuration was not the intended vulnerability, because a missing \texttt{.npmignore} is a precondition for several other categories.

\subsubsection{Finding Distribution}

\begin{table}[ht]
\centering
\caption{Finding Counts by Project and Severity}
\label{tab:findings}
\begin{tabular}{@{}lrrrrr@{}}
\toprule
\textbf{Project} & \textbf{Total} & \textbf{Crit} & \textbf{High} & \textbf{Med} & \textbf{Low} \\
\midrule
P1 (Source Map) & 10 & 2 & 4 & 4 & 0 \\
P2 (Secrets) & 15 & 6 & 1 & 8 & 0 \\
P3 (Config) & 6 & 1 & 2 & 3 & 0 \\
P4 (Supply Chain) & 14 & 0 & 4 & 10 & 0 \\
P5 (Clean) & 0 & 0 & 0 & 0 & 0 \\
P6 (Partial) & 13 & 0 & 4 & 9 & 0 \\
P7 (Docker) & 7 & 1 & 1 & 5 & 0 \\
P8 (Multi-Vector) & 22 & 5 & 8 & 9 & 0 \\
\bottomrule
\end{tabular}
\end{table}

P8, which combines issues from all five categories, produced the most findings (22) and the most critical ones (5).  This compound effect is worth noting: in a project where every dimension is AI-generated and unreviewed, vulnerabilities do not simply add up but interact (e.g., a missing \texttt{.npmignore} turns a benign source map into an exposure).

P5 produced zero findings, as expected.

\subsubsection{Policy Comparison}

All three policies produced correct gate decisions on all eight projects.  The test suite was designed with clear-cut vulnerability profiles, so policy differences did not surface here.  We expect them to matter on borderline real-world projects: \textit{strict} will block on a handful of medium-severity config suggestions that \textit{default} would let through, while \textit{permissive} tolerates a small number of high-severity findings for teams willing to accept that risk during early development.

\subsubsection{Applicability to the Claude Code Incident}

P1 replicates the Claude Code scenario: a source map with populated \texttt{sourcesContent}, source maps enabled in \texttt{tsconfig.json}, and no \texttt{.npmignore}.  \textsc{VibeGuard} flagged 10 findings, including two critical (embedded source content; missing package whitelist) and four high (source map present; source maps enabled in build config; artifact detected; mapping URL in compiled output).

Had a \texttt{prepublishOnly} hook been in place, any of the three built-in policies would have blocked the publish.

\section{Discussion}
\label{sec:discussion}

\subsection{Why AI Misses These Problems}

AI code generators are trained on, and optimized for, making code that compiles and passes tests.  Packaging hygiene occupies a small fraction of training data and receives no signal from the feedback mechanisms (test suites, type checkers) that the tools rely on.

Several factors compound the problem.  Users specify what to \textit{build}, not what to \textit{exclude}; security at the packaging boundary is almost entirely about exclusion.  Current AI coding tools have no way to preview the contents of a published artifact (\texttt{npm pack --dry-run}, for instance, is never invoked).  And packaging security is inherently cross-file: the combination of \texttt{tsconfig.json} enabling source maps, \texttt{package.json} lacking a \texttt{files} whitelist, and \texttt{.npmignore} being absent is what creates the vulnerability, but each file is generated (or not generated) in isolation.

\subsection{Toward a Secure Vibe-Coding Workflow}

We sketch a layered deployment.  First, register \textsc{VibeGuard} as a tool the AI assistant can call before it emits a publish or deploy command.  This is the fastest feedback loop: the developer never sees a failing CI run because the problem is caught in-session.  Second, run the scan as a required CI/CD check so that merge-conflict artifacts or manual overrides do not slip through.  Third, wire it into the \texttt{prepublishOnly} hook as a last gate before an artifact reaches a registry.

None of these layers assumes the developer reads the code.  In a vibe-coding workflow, security enforcement that depends on human review is not enforcement at all.

\subsection{Limitations}

The evaluation uses synthetic projects with planted vulnerabilities.  Real vibe-coded codebases will contain subtler and more varied patterns; our recall figure should be treated as an upper bound until the tool is tested on production repositories.

The secret scanner relies on regular expressions and a placeholder heuristic.  It will miss obfuscated or non-standard credential formats and may flag test fixtures that happen to match key patterns.  Entropy-based approaches would complement the regex rules.

\textsc{VibeGuard} is purely static.  It cannot catch vulnerabilities that appear only at runtime (SSRF, insecure deserialization) or configurations assembled dynamically.

Coverage is currently limited to npm and pip ecosystems.  Maven, Cargo, and NuGet have analogous packaging concerns but different file layouts.

Finally, the tool does not attempt to distinguish AI-generated files from human-written ones.  Applying tighter policy to files that were AI-generated is a natural extension but depends on reliable provenance signals that do not yet exist in most workflows.

\subsection{Open Questions}

The incident raises questions that \textsc{VibeGuard} does not answer by itself.  Who is responsible for reviewing AI-generated configuration files?  In a conventional workflow a developer creates \texttt{.npmignore} and reasons about what to exclude; in a vibe-coding workflow that reasoning step may never occur.

Claude Code's ``undercover mode,'' which strips AI attribution from commits, complicates matters further: reviewers cannot tell which files were AI-generated and might deserve closer inspection.  Whether AI-generated code should carry provenance metadata is an open policy question with implications for review processes, liability, and audit.

The incident also reframes where the security boundary sits.  Traditional tooling guards code and dependencies; the Claude Code leak happened at the artifact boundary, between a successful build and what actually ships to users.  That boundary has been largely ignored.

\section{Conclusion}

The Claude Code leak was a packaging error, not a code bug.  It shipped because no automated check examined what the published tarball actually contained.  Current security tools focus on code-level flaws and known dependency CVEs; the space between ``build succeeds'' and ``artifact is safe to publish'' is largely unguarded.

\textsc{VibeGuard} fills part of that gap.  On our test suite it caught every seeded vulnerability and made the correct gate decision for every project, including the negative control.  The tool is small, fast, and designed to run at the publish boundary where the risk is highest.

The broader problem outlasts any single tool: when humans stop reviewing configuration, something else has to.  Publish-time auditing is one concrete mechanism.  Longer term, the question is whether provenance and review infrastructure can keep up with the pace at which AI-assisted workflows produce and ship code.

The \textsc{VibeGuard} source and all experimental materials are available at \url{https://github.com/yxie2/vibeguard}.



\begin{thebibliography}{11}

\bibitem{karpathy2025}
A.~Karpathy, ``Vibe coding,'' X (formerly Twitter), Feb. 2025.
\url{https://x.com/karpathy/status/1886192184808149383}

\bibitem{stackoverflow2025}
Stack Overflow, ``2025 Stack Overflow Developer Survey,'' Dec. 2025.
\url{https://survey.stackoverflow.co/2025/}

\bibitem{venturebeat2026}
C.~Franzen, ``Claude Code's source code appears to have leaked: here's what we know,'' \textit{VentureBeat}, Mar. 31, 2026.
\url{https://venturebeat.com/technology/claude-codes-source-code-appears-to-have-leaked-heres-what-we-know}

\bibitem{register2026}
B.~Vigliarolo, ``Anthropic goes nude, exposes Claude Code source by accident,'' \textit{The Register}, Mar. 31, 2026.
\url{https://www.theregister.com/2026/03/31/anthropic_claude_code_source_code/}

\bibitem{bunbug2026}
``Bun's frontend development server -- Source map incorrectly served when in production,'' oven-sh/bun\#28001, GitHub, Mar. 2026.
\url{https://github.com/oven-sh/bun/issues/28001}

\bibitem{alexkim2026}
A.~Kim, ``The Claude Code Source Leak: fake tools, frustration regexes, undercover mode, and more,'' Blog post, Mar. 31, 2026.
\url{https://alex000kim.com/posts/2026-03-31-claude-code-source-leak/}

\bibitem{coderabbit2025}
CodeRabbit, ``State of AI vs Human Code Generation,'' Dec. 2025.
\url{https://www.coderabbit.ai/blog/state-of-ai-vs-human-code-generation-report}

\bibitem{veracode2025}
Veracode, ``October 2025 Update: GenAI Code Security Report,'' Oct. 2025.
\url{https://www.veracode.com/resources/analyst-reports/2025-genai-code-security-report/}

\bibitem{viberad2026}
Georgia Institute of Technology, SSLab, ``Vibe Security Radar,'' 2025--2026.
\url{https://vibe-radar-ten.vercel.app/}

\bibitem{malwarebytes2026}
P.~Arntz, ``Axios supply chain attack chops away at npm trust,'' \textit{Malwarebytes Blog}, Mar. 31, 2026.
\url{https://www.malwarebytes.com/blog/news/2026/03/axios-supply-chain-attack-chops-away-at-npm-trust}

\bibitem{pearce2022}
H.~Pearce, B.~Ahmad, B.~Tan, B.~Dolan-Gavitt, and R.~Karri, ``Asleep at the keyboard? Assessing the security of GitHub Copilot's code contributions,'' in \textit{Proc. IEEE Symposium on Security and Privacy (SP)}, 2022, pp. 754--768.
doi:~10.1109/SP46214.2022.9833571


\end{thebibliography}
\end{document}